\documentclass[10pt,conference]{IEEEtran}

\makeatletter
\def\ps@headings{%
\def\@oddhead{\mbox{}\scriptsize\rightmark \hfil \thepage}%
\def\@evenhead{\scriptsize\thepage \hfil \leftmark\mbox{}}%
\def\@oddfoot{}%
\def\@evenfoot{}}
\makeatother
\usepackage{amsfonts}
\usepackage{mathrsfs}
\usepackage{amsfonts}
\usepackage{amssymb}
\usepackage{graphicx}
\usepackage{epsfig}
\usepackage{epstopdf}
\usepackage{psfrag}
\usepackage{amsmath}
\usepackage{array}
\usepackage{subfigure}
\usepackage{cite,graphicx,amsmath,amssymb,color}
\usepackage{anysize}
\usepackage{algorithmic}
\usepackage{algorithm}

\usepackage{algorithm}
\usepackage{algorithmic}
\usepackage{stmaryrd}
\usepackage{multirow}
\usepackage{subfig}
\usepackage{graphicx,times,amsmath} 

\usepackage{url}
 \usepackage{enumerate}

\marginsize{13mm}{13mm}{19mm}{43mm}

\IEEEoverridecommandlockouts

\newtheorem{theorem}{Theorem}
\newtheorem{lemma}{Lemma}

\newtheorem{problem}{Problem}
\newtheorem{remark}{Remark}

\begin{document}
\bibliographystyle{IEEEtran}

\title{Energy-Efficient Resource Allocation for Multi-User Mobile Edge Computing}
\author{Junfeng Guo, \quad  Zhaozhe Song, \quad Ying Cui\\
Department of Electronic Engineering, Shanghai Jiao Tong University, Shanghai, P. R. China}
\maketitle

\begin{abstract}
To increase mobile batteries' lifetime and improve quality of experience for computation-intensive and latency-sensitive applications, mobile edge computing has received significant interest. Designing energy-efficient mobile edge computing systems requires joint optimization of communication and computation resources. In this paper, we consider energy-efficient resource allocation for a multi-user mobile edge computing system. First, we establish on two computation-efficient models with negligible and non-negligible base station (BS) executing durations, respectively. Then, under each model, we formulate the overall weighted sum energy consumption minimization problem by optimally allocating communication and computation resources. The optimization problem for negligible BS executing duration is convex, and we obtain the optimal solution in closed-form to this problem. The optimization problem for non-negligible BS executing duration is NP-hard in general, and we obtain a sub-optimal solution with low-complexity to this problem, by connecting it to a three-stage flow-shop scheduling problem and wisely utilizing Johnson's algorithm. Finally, numerical results show that the proposed solutions outperform some baseline schemes. 
\end{abstract}


\section{Introduction}
With the support of on-device cameras and embedded sensors, new applications with advanced features, e.g., navigation, Augmented Reality, interactive online gaming and multi-media transformation like Speech2Text, have been developed. These applications are both computation-intensive and latency-sensitive. Mobile edge computing (MEC) is a promising technology providing an IT service environment and cloud-computing capabilities at the edge of the mobile network, within the Radio Access Network and in close proximity to mobile users to improve quality of experience. In an MEC system, computation tasks of mobile users are uploaded to the base station (BS) and executed at the MEC servers. Then, the computation results are transmitted back to the mobiles. With the drastic demands of applications with crucial computation and latency requirements, the finite battery lifetime and limited communication and computation resources pose challenges for designing future energy-efficient MEC systems\cite{ahmed2016survey,hu2015mobile}.

Designing of energy-efficient MEC requires the joint allocation of communication and computation resources among distributed mobiles and MEC servers at BSs. Emerging research toward this direction considers optimal resource allocation for various types of MEC systems~\cite{mao2016dynamic,liu2016delay,zhang2013energy,you2016energy0,you2016energy}.
For example, \cite{mao2016dynamic} and \cite{liu2016delay} consider  single-user MEC systems with one BS and multiple elastic tasks, and study optimal offloading control and resource allocation (such as CPU-cycle frequency allocation and power allocation) to minimize the average execution delay of all tasks under power constraints.  
In \cite{zhang2013energy} and \cite{you2016energy0}, the authors consider single-user MEC systems with one BS and a single inelastic task, and study optimal offloading control and resource allocation to minimize the total energy consumption under a hard deadline constraint. In particular, \cite{zhang2013energy} obtains an optimal threshold-based offloading policy and optimal CPU-cycle frequency and power allocation. In \cite{you2016energy0}, optimal CPU-cycle frequency allocation and time division between microwave power transfer and offloading are derived.  
Reference \cite{you2016energy} considers a multi-user MEC system with one BS and an inelastic task for each user, and studies optimal task splitting and resource allocation to minimize the weighted sum mobile energy consumption under a hard deadline constraint. It is shown in \cite{you2016energy} that the optimal task splitting policy is of a threshold-based structure. 


Note that all the aforementioned papers have the following limitations. (i) All these papers assume that the size of the computation result for each task is negligible, and fail to take account of the resource consumption for downloading the computation results back to mobiles. Thus, the obtained solutions are not suitable for the applications with large-size computation results, such as Augmented Reality, interactive online gaming and multi-media transformation. (ii) All these papers assume that BS executing duration is negligible, and hence do not consider the processing order of tasks for offloading and executing. This assumption is not reasonable when multiple users simultaneously offload  computation-intensive and latency-sensitive tasks to the same BS for executing. 
Note that offloading and executing operations can be conducted in parallel, and the processing order and total completion time of tasks greatly affect the energy consumption. In summary, further studies are required to design energy-efficient multi-user MEC systems to ultimately provide satisfactory quality of experience for computation-intensive and latency-sensitive applications.

In this paper, we shall address the above issues. 
We consider energy-efficient resource allocation for a multi-user MEC system with one BS of computing capability and multiple users each with an inelastic task.
First, we propose a more practical task model, specifying each task using three parameters, i.e., size of the task before computation, workload and size of the computation result. 
We establish on two computation-offloading models with negligible and non-negligible BS executing durations, respectively. Under each model, we formulate the overall weighted sum energy consumption minimization problem by optimally allocating communication and computation resources. The optimization problem for negligible BS executing duration is convex, and we obtain the optimal uploading and downloading duration allocation for each task in closed-form. The optimization problem for non-negligible BS executing duration is NP-hard in general, and we obtain a sub-optimal processing order for all tasks and the optimal uploading and downloading duration allocation for each task under this order, by connecting the problem to a three-stage flow-shop scheduling problem and wisely utilizing Johnson's algorithm. We show that the sub-optimal solution has promising performance and low-complexity. Finally, numerical results show that the proposed solutions outperform some baseline schemes. 
 
%

\section{System Model}
\begin{figure}[H]
\begin{center}
  \subfigure[\small{Multi-user MEC system.}]
  {\resizebox{4.4cm}{!}{\includegraphics{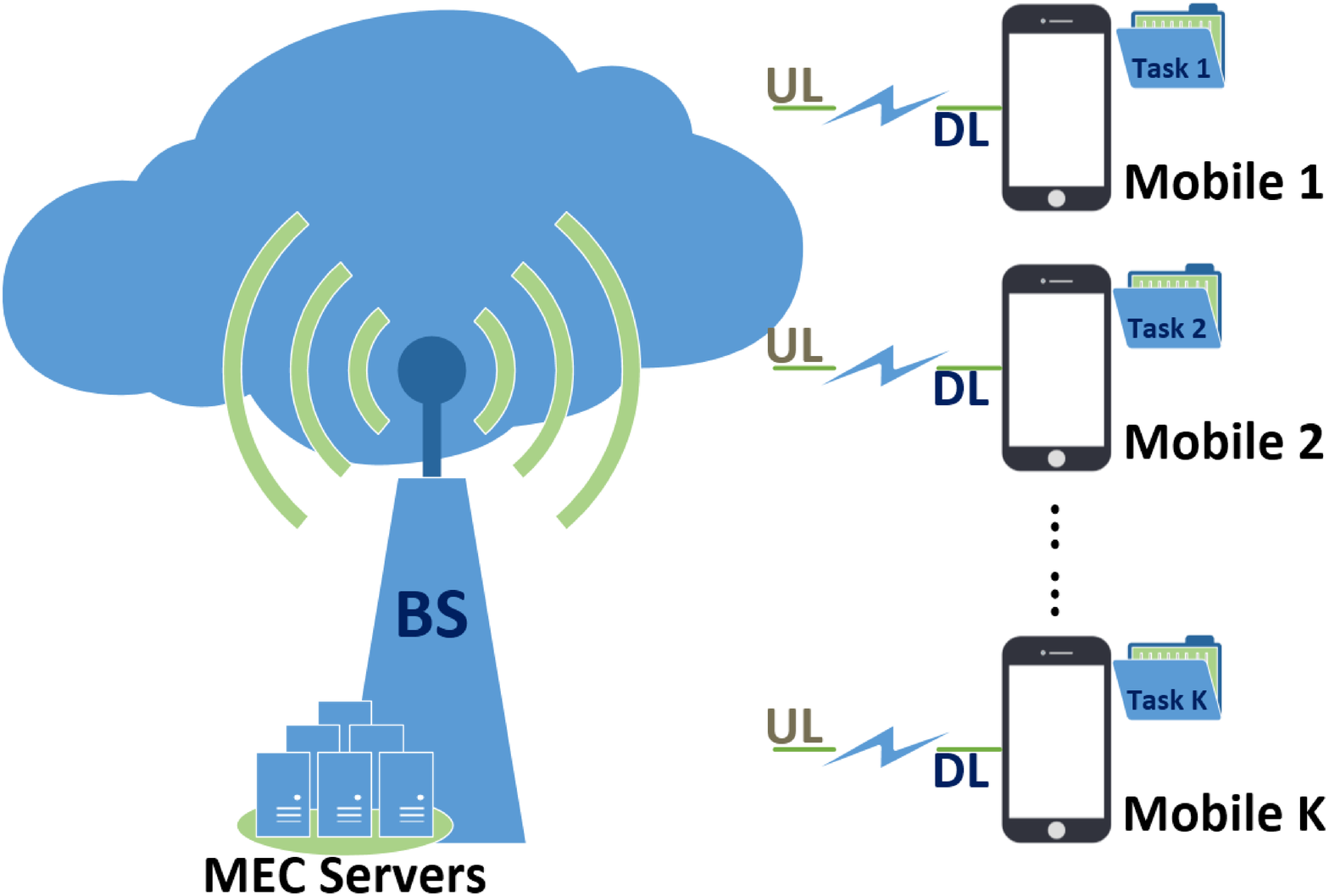}}}
  \subfigure[\small{Three operations.}]
  {\resizebox{4.4cm}{!}{\includegraphics{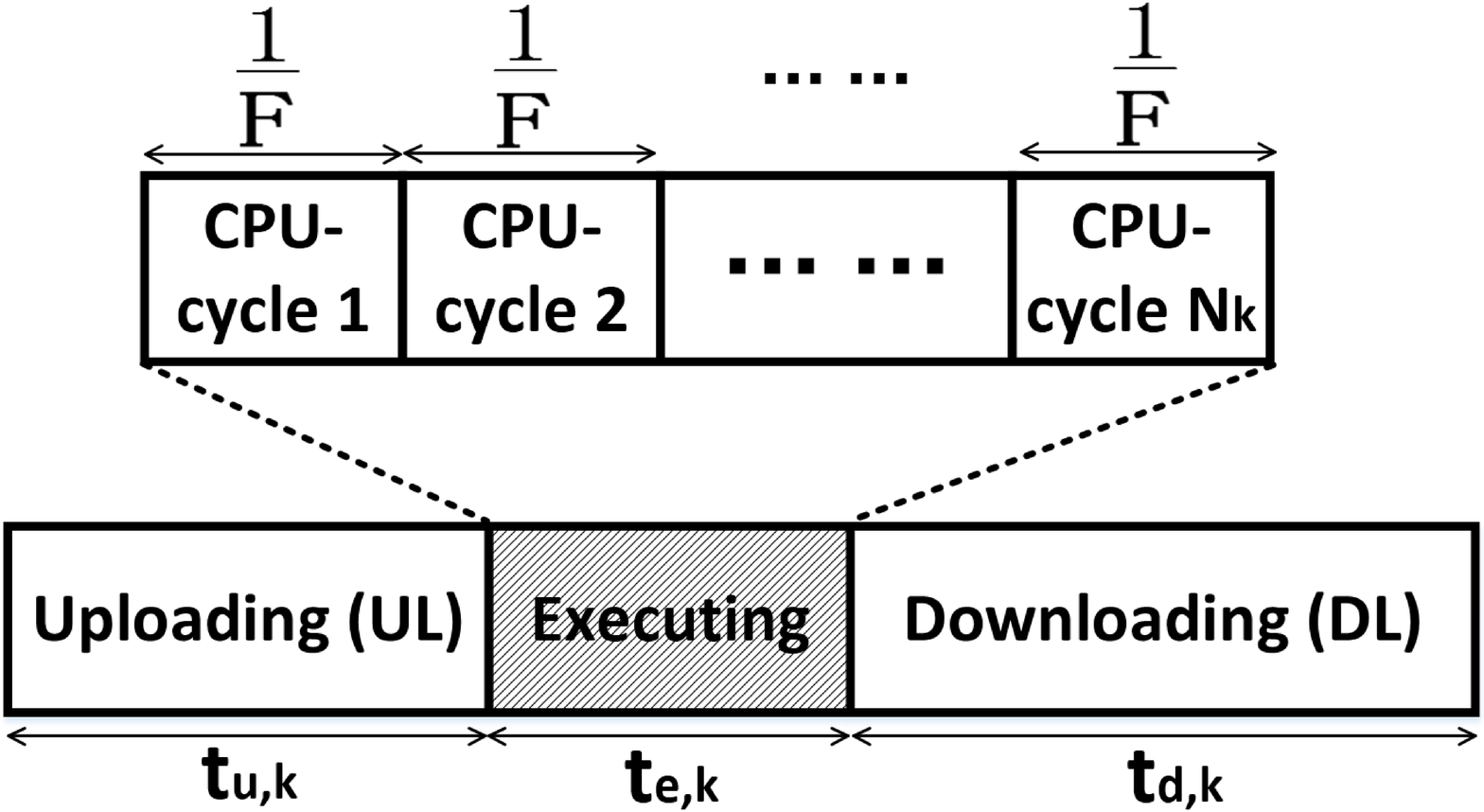}}}
  \end{center}
         \caption{\small{ System model. }}
\label{system_model}
\end{figure}

As illustrated in Fig.~\ref{system_model}, we consider a multi-user MEC system consisting of one single-antenna base station (BS) and $K$ single-antenna mobiles, denoted by set $\mathcal K \triangleq \{1, 2, ...,K\}$. The BS has powerful computing capability by running IT based servers of a constant CPU-cycle frequency (in number of CPU-cycles per second) at the network edge. Each mobile has a computation-intensive and latency-sensitive (computation) task, which is offloaded to the BS for executing.\footnote{We assume that all tasks have to be executed at the BS due to crucial computation and latency requirements. The optimization results obtained in this paper can be used to study optimal offloading control (to determine the sets of tasks executed locally and offloaded to the BS for executing) by considering a discrete optimization problem. This is beyond the scope of this paper}

We first propose a more practical computation task model.
The computation task at mobile $k \in \mathcal K$, referred to as task $k$, is characterized by three parameters, i.e., the size of the task before computation $L_{u,k} > 0$ (in bits), workload $N_k > 0$ (in number of CPU-cycles), and the size of the computation result $L_{d,k} > 0$ (in bits). The computation of each task $k$ has to be accomplished within $T$ seconds. 

\begin{remark}[Task Model]
Note that this computation task model generalizes those in \cite{mao2016dynamic,liu2016delay,zhang2013energy,you2016energy0,you2016energy} in the sense that the size of the computation result is taken into consideration.
\end{remark}

Offloading task $k$ to the BS for executing comprises three sequential operations: 1) uploading task $k$ of $L_{u,k}$ bits from mobile $k$ to the BS; 2) executing task $k$ at the BS (which requires $N_k$ BS CPU-cycles); 3) downloading the computation result of $L_{d,k}$ bits from the BS to mobile $k$. Let $t_{u,k} \geq 0$, $t_{e,k} \geq 0$ and $t_{d,k} \geq 0$ denote the uploading, executing and downloading durations in the three operations, respectively. Let $F > 0$ denote the fixed CPU-cycle frequency at the BS.
The BS executing duration (in seconds) is given by:
\begin{equation}\label{executing_time}
t_{e,k} = \frac{N_k}{F}.
\end{equation}
Since $F$ is usually large, $t_{e,k}$ is small.
In the following, we first consider a computation-offloading model with negligible BS executing duration (i.e., $t_{e,k} \approx 0$). In Section~\ref{section4}, we will consider a computation-offloading model with non-negligible BS executing duration. We consider Time Division Multiple Access (TDMA) with Time-Division Duplexing (TDD) operation. The processing order of $K$ tasks does not matter when the BS executing duration is negligible, since the total completion time is always the sum of the uploading and downloading durations for all tasks. Thus, the uploading and downloading duration allocation satisfies:
\begin{align}
& 0 \leq t_{u,k} \leq T,~ k \in \mathcal K, \label{time_constraint_1} \\
& 0 \leq t_{d,k} \leq T,~ k \in \mathcal K, \label{time_constraint_2} \\
& \sum_{k \in \mathcal K}(t_{u,k}+t_{d,k}) \leq T. \label{time_constraint_3}
\end{align}

Similar to \cite{you2016energy} and \cite{chandrakasan1992low}, we consider low CPU voltage at the BS, and model the energy consumption for BS executing as follows.\footnote{The circuit power is omitted here for simplicity but can be accounted for by adding a constant \cite{you2016energy,chandrakasan1992low}.} At the BS, the amount of energy consumption for computation in a single CPU-cycle with frequency $F$ is $\mu F^2$, where $\mu$ is a constant factor determined by the switched capacitance at the MEC servers. Then, the energy consumption for executing task $k$ at the BS is given by:
\begin{equation}
E_{e,k} \triangleq \mu N_k F^2.
\end{equation}

We now introduce the energy consumption model for task uploading and downloading operations.
Let $h_k$ denote the channel power gain for mobile $k$ which is assumed to be constant during the $T$ seconds. Let $p_k$ denote the transmission power of mobile $k$ for uploading task $k$. Then, the achievable transmission rate (in bit/s) for uploading task $k$ is given by:
\begin{align}\label{transmission_rate}
r_k = B \log_2 \left(1 + \frac{p_k |h_k|^2}{n_0}\right),
\end{align}
where $B$ and $n_0$ are the bandwidth (in Hz) and the variance of complex white Gaussian channel noise, respectively. On the other hand, the transmission rate for uploading task $k$ is fixed as $r_k = L_{u,k}/t_{u,k}$, since this is the most energy-efficient transmission method for transmitting $L_{u,k}$ bits in $t_{u,k}$ seconds. 
Define $g(x) \triangleq n_0 \left(2^{\frac{x}{B}} - 1\right)$. Then, we have $p_k = \frac{1}{|h_k|^2}g\left(\frac{L_{u,k}}{t_{u,k}}\right)$. Thus, the energy consumption at mobile $k$ for uploading task $k$ to the BS is given by:
\begin{equation}
E_{u,k}(t_{u,k}) \triangleq p_k t_{u,k} = \frac{t_{u,k}}{|h_k|^2} g\left(\frac{L_{u,k}}{t_{u,k}}\right).
\end{equation}
Similarly, the energy consumption at the BS for transmitting the computation result of task $k$ to mobile $k$ is given by:
\begin{equation}
E_{d,k}(t_{d,k}) \triangleq \frac{t_{d,k}}{|h_k|^2}g\left(\frac{L_{d,k}}{t_{d,k}}\right).
\end{equation}
Thus, the energy consumption at the BS for executing and transmitting task $k$ is given by: 
\begin{equation}
E_{BS,k}(t_{d,k}) = E_{e,k} + E_{d,k}(t_{d,k}).
\end{equation}
The weighted sum energy consumption for executing task $k$ by offloading to the BS is given by:
\begin{equation} 
E_{k}(t_{u,k},t_{d,k}) = E_{u,k}(t_{u,k}) + \beta E_{BS,k}(t_{d,k}),
\end{equation}
where $\beta \geq 0$ is the corresponding weight factor.
Therefore, the overall weighted sum energy consumption for executing the $K$ tasks by offloading to the BS is given by:
\begin{equation}\label{offload_energy}
E (\mathbf t_u,\mathbf t_d) = \sum_{k \in \mathcal K}  E_{k}(t_{u,k},t_{d,k}),
\end{equation}
where $\mathbf t_u \triangleq (t_{u,k})_{k \in \mathcal K}$ and $\mathbf t_d \triangleq (t_{d,k})_{k \in \mathcal K}$.

\begin{remark}[Energy Consumption Model]
Note that the computation-offloading model with negligible BS executing duration considered in this paper is a generalization of those in \cite{mao2016dynamic,liu2016delay,zhang2013energy,you2016energy0,you2016energy} in the sense that task downloading and the corresponding resource consumption are considered.
\end{remark}

\section{Multi-user MEC with Negligible BS Executing Duration}\label{section3}
In this section, we first formulate the energy minimization problem for the multi-user MEC system with negligible BS executing duration. Then, we obtain the optimal solution.


\subsection{Problem Formulation}
We would like to minimize the overall weighted sum energy consumption under the uploading and downloading duration allocation constraints. Specifically, we have the following optimization problem.

\begin{problem}[Negligible BS Executing Duration] \label{offloading_problem}
\begin{align*}
E^* \; \triangleq \; \min_{\mathbf t_u,\mathbf t_d} \;\; & E (\mathbf t_u,\mathbf t_d) \\
s.t.\;\; &\eqref{time_constraint_1},\eqref{time_constraint_2},\eqref{time_constraint_3}.
\end{align*}
\end{problem}

\subsection{Optimal Solution}
We can easily verify that Problem~\ref{offloading_problem} is convex and Slater's condition is satisfied, implying that strong duality holds. Thus, Problem~\ref{offloading_problem} can be solved using KKT conditions.

\begin{lemma}[Optimal Solution to Problem \ref{offloading_problem}] \label{solve_t}
The optimal solution $(\mathbf t_u^*,\mathbf t_d^*)$ to Problem~\ref{offloading_problem} is given by:
\begin{align}
    \begin{cases}
    t_{u,k}^* \; = \; \frac{L_{u,k} \ln2}{B\left(W\left(\frac{\lambda^* |h_k|^2 - n_0}{n_0 e}\right) + 1 \right)~}   \\
    t_{d,k}^* \; = \; \frac{\beta L_{d,k} \ln2}{B\left(W\left(\frac{\lambda^* |h_k|^2 - n_0}{n_0 e}\right) + 1 \right)~}   \\
    \end{cases}
    ,~ k \in \mathcal K,
\end{align}
where $W(\cdot)$ denotes the Lambert function and $\lambda^*$ satisfies: 
\begin{equation} \label{active_constraint}
\sum_{k \in \mathcal K} \; \frac{\left(L_{u,k}+\beta L_{d,k}\right) \ln2}{B \left(W\left(\frac{\lambda^* |h_k|^2 - n_0}{n_0 e}\right) + 1 \right)~} \; = \; T.
\end{equation}

Note that, $\lambda^*$ in \eqref{active_constraint} can be easily obtained using the bisection method. Thus, using Lemma \ref{solve_t}, we can compute $(\mathbf t_u^*,\mathbf t_d^*)$ efficiently.

\end{lemma}

\begin{remark}[Interpretation of Lemma \ref{offloading_problem}]
The optimal solution adapts to the operations and the channel conditions.
In particular, for given $\lambda^*$, the uploading and downloading durations of a task increase with the size of the task and the size of the computation result, respectively, and both decrease with the channel power gain.
\end{remark}

\section{Multi-user MEC with Non-negligible BS Executing Duration}\label{section4}
In this section, we consider a more practical scenario where the BS executing duration is non-negligible. 
We first elaborate on the computation-offloading model in this scenario. Then, we formulate the energy minimization problem for the multi-user MEC system with BS executing duration. Finally, we characterize the optimal solution and propose a low-complexity sub-optimal solution with promising performance.

\subsection{Computation-Offloading Model with Non-negligible BS Executing Duration}

\begin{figure}[t]
\begin{center}
  \subfigure[\small{Arbitrary processing sequences and starting time.}]
  {\resizebox{7cm}{!}{\includegraphics{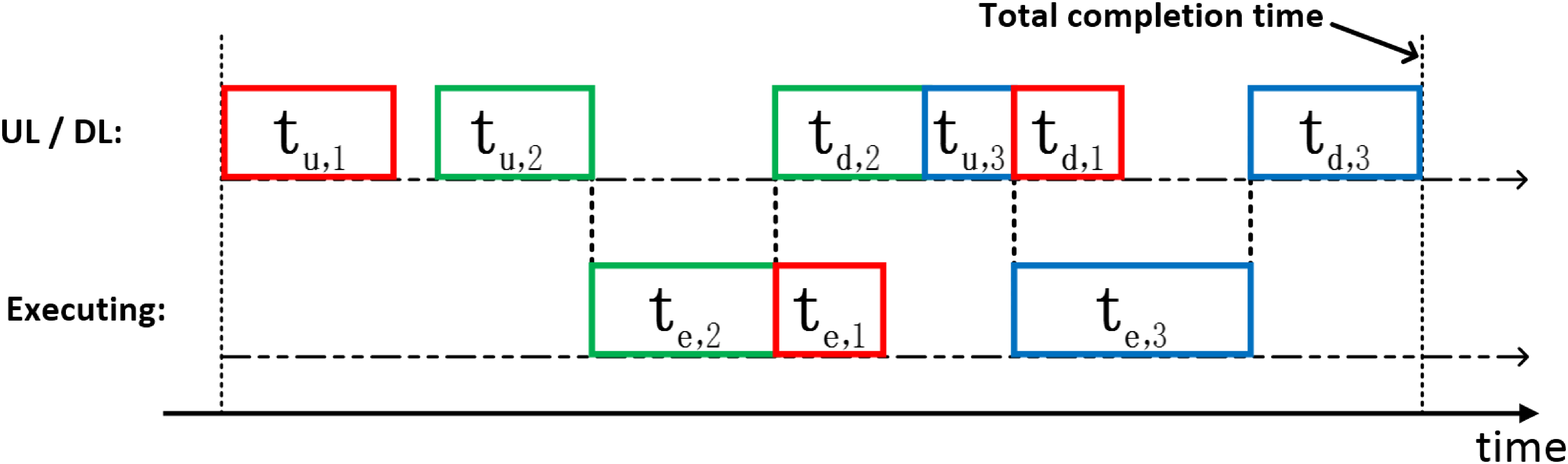}}}
  \subfigure[\small{Same processing sequence for uploading,executing and downloading operation.}]
  {\resizebox{7cm}{!}{\includegraphics{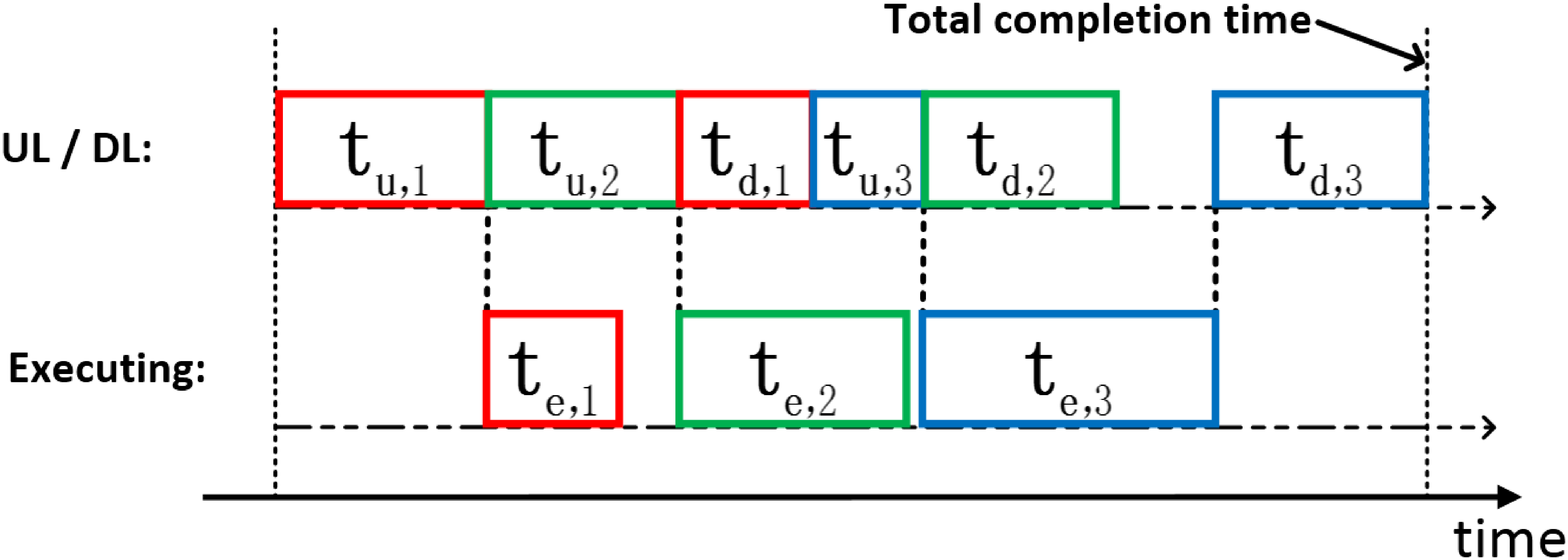}}}
  \subfigure[\small{Completing the uploading operations of all tasks before starting the downloading operation of any task.}]
  {\resizebox{7cm}{!}{\includegraphics{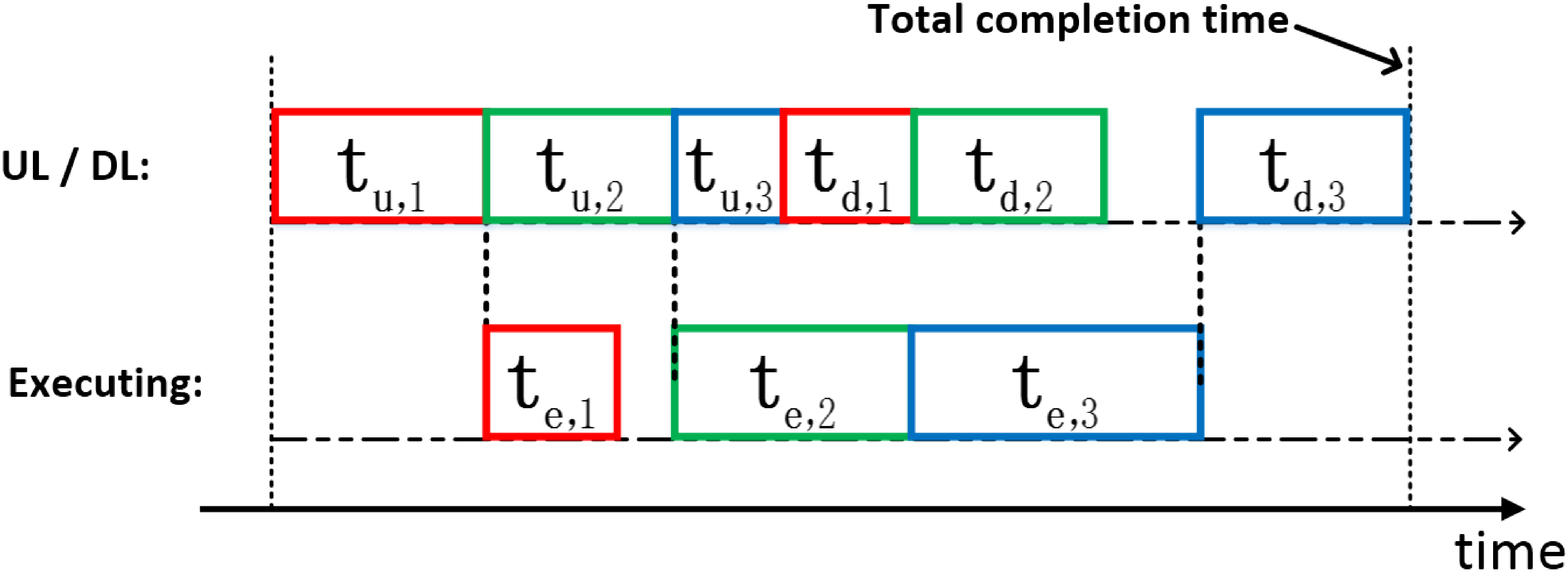}}}
  \end{center}
         \caption{\small{ Illustration example of three operations of all tasks at the BS, i.e., $\mathcal K = \{1,2,3\}$. For each task, the required duration of each operation is represented by the length of the corresponding rectangle.}}
\label{example}
\end{figure}

In this part, task executing duration at the BS is considered. In addition, note that the execution of one task and the transmission of another task can be conducted at the same time. These make the computation-offloading model with non-negligible BS executing duration sufficiently different from the one without BS executing duration, as illustrated in Fig.~\ref{example}(a). In the following, we introduce new notations and constraints to specify this model.

Let $s_{u,k}$, $s_{e,k}$ and $s_{d,k}$ denote the starting times for uploading, executing and downloading task $k$, respectively. Let $c_{u,k}$, $c_{e,k}$ and $c_{d,k}$ denote the completion times for uploading, executing and downloading task $k$, respectively. As each of the three operations cannot be interrupted, we first have the following constraints:
\begin{align}
	\begin{cases}\label{def_constraint}
		s_{u,k} + t_{u,k} = c_{u,k} \\
	    s_{e,k} + t_{e,k} = c_{e,k} \\
        s_{d,k} + t_{d,k} = c_{d,k} \\
    \end{cases}
    ,~ k \in \mathcal K.
\end{align}
To ensure that the uploading, executing and downloading operations of task $k$ are conducted sequentially, we require:
\begin{align}
	\begin{cases}\label{user_constraint}
	    s_{u,k} \;\ge\; 0 \\
	    s_{e,k} \;\ge\; c_{u,k} \\
        s_{d,k} \;\ge\; c_{e,k} \\
    \end{cases}
    ,~ k \in \mathcal K.
\end{align}
To ensure that the downloading operation of task $k$ can be completed before deadline $T$, we have:
\begin{align}\label{time_constraint}
      c_{d,k} \;\leq\; T,~ k \in \mathcal K.
\end{align}

Based on $s_{u,k}$, $s_{e,k}$ and $s_{d,k}$ for all $k \in \mathcal{K}$ (or $c_{u,k}$, $c_{e,k}$ and $c_{d,k}$ for all $k\in \mathcal{K}$), we can obtain three orders (sequences) for uploading, executing, and downloading of the $K$ tasks, respectively. Following the proof of Lemma $3$ in \cite{johnson1954optimal}, we can show that the three sequences can be made the same without increasing the total completion time for processing all tasks, as illustrated in Fig.~\ref{example}(b). Thus, without loss of generality, we consider the same sequence for uploading, executing and downloading operations for the $K$ tasks, denoted by $\mathbf S \in \mathbb S$, where $\mathbb{S}$ denotes the set of the $K !$ different permutations of all tasks in $\mathcal K$. We let subscript $[k]$ denote the task index at position $k$ in sequence $\mathbf S$.
From Lemma $1$ in \cite{mathes19992}, we know that, completing the uploading operations of all $K$ tasks before starting the downloading operation of any task will not increase the total completion time, as illustrated in Fig.~\ref{example}(c). To ensure that at any time, there are at most one task under execution and at most one task under transmission, we have the following constraints:
\begin{align}
    &
    \begin{cases} \label{sequential_constraint}
        s_{u,[k]}  \;\ge\; c_{u,[k-1]} \\
        s_{e,[k]}  \;\ge\; c_{e,[k-1]} \\
        s_{d,[k]}  \;\ge\; c_{d,[k-1]} \\
    \end{cases}
    ,~ k = 2, 3, ... , K,\\
    &~~~ s_{d,[1]} \;\ge\;c_{u,[K]}.\label{nonparallel}
\end{align}  

The energy consumption model remains the same as that in Section \ref{section4}.

\begin{remark}[Non-negligible BS Executing Duration]
Note that the computation-offloading model further generalizes with non-negligible BS executing duration in Section~\ref{section3} in the sense that task offloading and the corresponding resource consumption are considered. Under this model, offloading and executing operations can be conducted in parallel, and the processing order and total completion time of all tasks greatly affect the energy consumption. 
\end{remark}

\subsection{Problem Formulation}
We would like to minimize the overall weighted sum energy consumption for the multi-user MEC with non-negligible BS executing duration under the uploading and downloading duration allocation constraints. Specifically, we have the following optimization problem.
\begin{problem}[Non-negligible BS Executing Duration] \label{C_original}
\begin{align*}
\min_{{\mathbf S \in \mathbb{S}},{\mathbf s_u},{\mathbf s_e},{\mathbf s_d},\mathbf t_u,\mathbf t_d} \; & E (\mathbf t_u,\mathbf t_d)  \\
\;\;\;\;s.t.\;\;\;\;\;\;\;&
            \eqref{time_constraint_1},\eqref{time_constraint_2},\eqref{def_constraint},\eqref{user_constraint},\eqref{time_constraint},\eqref{sequential_constraint},\eqref{nonparallel},
\end{align*}
where $\mathbf s_u \triangleq (s_{u,k})_{k \in \mathcal K}$, $\mathbf s_e \triangleq (s_{e,k})_{k \in \mathcal K}$ and $\mathbf s_d \triangleq (s_{d,k})_{k \in \mathcal K}$.
\end{problem}

Problem \ref{C_original} is a mixed discrete-continuous optimization problem with two main challenges. One is the choice of the operation sequence selection (discrete variable), and the other is the choice of the uploading and downloading duration allocation (continuous variables). We thus propose an equivalent alternative formulation of Problem~\ref{C_original} which naturally subdivides Problem~\ref{C_original} according to these two aspects.

\begin{problem}[Sequence Selection] \label{C_searchS}
\begin{align*}
\widetilde{E}^{*} \; \triangleq \; \min_{\mathbf{S}}  \;\; & E_{\mathrm{seq}}^*(\mathbf{S}) \\
s.t.\;\; & \mathbf{S} \in \mathbb{S} .
\end{align*}
Let $\mathbf S^*$ denote the optimal solution. $E_{\mathrm{seq}}^*(\mathbf{S})$ is given by the following sub-problem.
\end{problem}
\begin{problem}[Duration Allocation] \label{C_offload}
For any $\mathbf{S} \in \mathbb{S}$, we have
\begin{align*}
E_{\mathrm{seq}}^*(\mathbf{S}) \; \triangleq \;  & \min_{{\mathbf{s_u}},{\mathbf{s_e}},{\mathbf{s_d}},\mathbf t_u,\mathbf t_d} \; E (\mathbf t_u,\mathbf t_d) \\
\;\;\;\;&\;\;\;\;\;\;\;s.t.\;\;\;\;\;\;\;\; \eqref{time_constraint_1},\eqref{time_constraint_2},\eqref{def_constraint},\eqref{user_constraint},\eqref{time_constraint},\eqref{sequential_constraint},\eqref{nonparallel}.
\end{align*}
\end{problem}

\subsection{Optimal Solution}
First, we obtain an optimal solution to Problem~\ref{C_offload} for given $\mathbf{S} \in \mathbb{S}$. Problem~\ref{C_offload} is a convex optimization problem. The number of variables in Problem~\ref{C_offload} is $5K$, which is huge for large $K$. Thus, the complexity for solving Problem~\ref{C_offload} is very high when $K$ is large. We would like to reduce the computational complexity. By exploiting structural properties of the constraints in Problem~\ref{C_offload}, we first obtain the minimum total completion time for all tasks under given $\mathbf S \in \mathbb{S},\mathbf t_u$ and $\mathbf t_d$, denoted by $T_F(\mathbf S, \mathbf t_u,\mathbf t_d)$.

\begin{lemma}[Minimum Total Completion Time] \label{cal_t}
For given $\mathbf S \in \mathbb{S}, \mathbf t_u$ and  $\mathbf t_d$, the minimum total completion time is given by \eqref{cal_t2} (at the top of the next page).

\begin{figure*}[!t]
\small{\begin{align} 
 T_F(\mathbf S, \mathbf t_u,\mathbf t_d) \;& = 
 \max {\left\{\max_{1 \leq i \leq j \leq K} \left(\left(\sum_{k=1}^{j}t_{e,[k]} - \sum_{k=1}^{j-1} t_{d,[k]}\right) + \left(\sum_{k=1}^{i}t_{u,[k]} - \sum_{k=1}^{i-1}t_{e,[k]} \right)\right),\sum_{k=1}^{K}t_{u,[k]}\right\}}+\sum_{k=1}^{K}t_{d,[k]}.\label{cal_t2}
 \\
\widetilde T_F(\mathbf S, \mathbf t_u,\mathbf t_d) \;& = 
\max_{1 \leq i \leq j \leq K} \left(\left(\sum_{k=1}^{j}t_{e,[k]} - \sum_{k=1}^{j-1} t_{d,[k]}\right) + \left(\sum_{k=1}^{i}t_{u,[k]} - \sum_{k=1}^{i-1}t_{e,[k]} \right)\right)+\sum_{k=1}^{K}t_{d,[k]}.\label{T_F}
\end{align}} \hrulefill
\end{figure*}

\end{lemma}

We now introduce another convex optimization problem, by replacing the constraints in Problem~\ref{C_offload} with a deadline constraint on the minimum total completion time $T_F(\mathbf S, \mathbf t_u,\mathbf t_d)$ in \eqref{cal_t2}.

\begin{problem}[Equivalent Problem of Problem \ref{C_offload}] \label{C_transed}
For any $\mathbf{S} \in \mathbb{S}$, we have
\begin{align*}
E_{\mathrm{seq}}^*(\mathbf{S}) \; \triangleq \; 
\min_{\mathbf t_u,\mathbf t_d}\;\; &E (\mathbf t_u,\mathbf t_d) \\
s.t.\;\;    & T_F(\mathbf S, \mathbf t_u,\mathbf t_d) \le T,\\
            & \eqref{time_constraint_1},\eqref{time_constraint_2}.
\end{align*}
Let $(\mathbf t_u^*(\mathbf S),\mathbf t_d^*(\mathbf S))$ denote the optimal solution.
\end{problem}

\begin{theorem}[Relationship Between Problems~\ref{C_offload} and \ref{C_transed}] \label{transform_p}
Problem~\ref{C_offload} and Problem~\ref{C_transed} are equivalent.
\end{theorem}

Note that Problem~\ref{C_transed} is convex with $2K$ variables and can be solved more efficiently. Thus, for given $\mathbf{S} \in \mathbb{S}$, we solve Problem~\ref{C_transed} instead of Problem~\ref{C_offload} to obtain $E_{\mathrm{seq}}^*(\mathbf{S})$. Finally, we can solve Problem \ref{C_searchS} by evaluating all possible choices for $\mathbf S \in \mathbb{S}$ using exhaustive search.

\subsection{ Sub-optimal Solution}
Note that obtaining an optimal solution to Problem~\ref{C_searchS} requires solving Problem~\ref{C_transed} $K!$ times. The complexity is not acceptable when $K$ is large. In this part, by exploiting more structural properties, we obtain a low-complexity sub-optimal solution to Problem~\ref{C_searchS}.
Specifically, we connect Problem~\ref{C_searchS} to the conventional three-stage flow-shop scheduling problem \cite{johnson1954optimal} and solve it by utilizing Johnson's algorithm in \cite{johnson1954optimal}.
Obtaining the sub-optimal solution to Problem~\ref{C_searchS} only requires solving Problem~\ref{offloading_problem} once and solving Problem~\ref{C_transed} at most once.

First, we introduce some background on $M$-stage flow-shop scheduling problems. In an $M$-stage flow-shop scheduling problem, all tasks have to be processed on $M$ machines following the same machine order. Each task requires certain fixed processing time on a machine. The objective is to find a sequence for processing the tasks on each machine so that a given criterion is optimal. The criterion that is most commonly studied in the literature is the total completion time. When $M \geq 3$, an $M$-stage flow-shop scheduling problem is NP-hard in general. When $M = 3$, the three sequences for processing the tasks on the three machines can be set to be the same without losing optimality, and the optimal sequence can be obtained by Johnson's algorithm  in a special case\cite{johnson1954optimal}.

We now connect Problem \ref{C_searchS} to a three-stage flow-shop scheduling problem. First, we transform Problem~\ref{C_searchS} to an equivalent problem.
\begin{problem}[Equivalent Problem of Problem~\ref{C_searchS}] \label{tran}
\begin{align*}
{E}^{*} \; \triangleq \; \min_{\mathbf t_u,\mathbf t_d} \;\; &  E(\mathbf t_u,\mathbf t_d) \\
s.t.\;\; & T_F^*(\mathbf t_u,\mathbf t_d) \leq T \label{completion},
\\ 
         & \eqref{time_constraint_1},\eqref{time_constraint_2},
\end{align*}
Let $(\mathbf t_u^*,\mathbf t_d^*)$ denote the optimal solution. $T_F^*(\mathbf t_u,\mathbf t_d)$ is the optimal value of the following problem.
\end{problem}
\begin{problem}[Three-Stage Scheduling Problem] \label{schedule}
For any $\mathbf t_u$ and $\mathbf t_d$, we have
\begin{align*}
    T_F^*(\mathbf t_u,\mathbf t_d) \; \triangleq \min_{\mathbf{S} \in \mathbb{S}}\;\; & T_F(\mathbf S, \mathbf t_u,\mathbf t_d)\\
    s.t.\;\; & \eqref{def_constraint},\eqref{user_constraint},\eqref{time_constraint},\eqref{sequential_constraint},\eqref{nonparallel}.
\end{align*}

\end{problem}

By treating $\mathbf t_u$, $\mathbf t_e$ and ${\mathbf t_d}$ as the processing times for three separate machines (i.e., uploading machine, executing machine and downloading machine), Problem \ref{schedule} can be regarded as a three-stage flow-shop scheduling problem with an additional constraint in \eqref{nonparallel} (i.e., the uploading machine and downloading machine cannot operate at the same time). 
By relaxing the additional constraint in \eqref{nonparallel} and using the minimum total completion time (without the additional constraint), we can transform Problem~\ref{schedule} into a standard three-stage flow-shop scheduling problem\cite{johnson1954optimal}.
\begin{problem}[Three-Stage Flow-Shop Scheduling Problem] \label{flowshop}
For any $\mathbf t_u$ and $\mathbf t_d$, we have
\begin{align*}
    \widetilde T_F^*(\mathbf t_u,\mathbf t_d) \; \triangleq \; \min_{\mathbf{S} \in \mathbb{S}}\;\; & \widetilde T_F(\mathbf S, \mathbf t_u,\mathbf t_d)\\
    s.t.\;\;\; & \eqref{def_constraint},\eqref{user_constraint},\eqref{time_constraint},\eqref{sequential_constraint},
\end{align*}
where the minimum total completion time (without the additional constrain) $\widetilde T_F(\mathbf S, \mathbf t_u,\mathbf t_d)$ is given by \eqref{T_F} (at the top of next page) \cite{johnson1954optimal}.
Let $\mathbf{S}^*(\mathbf t_u,\mathbf t_d)$ denote an optimal solution. 
\end{problem}

It can be easily verified that Problem \ref{flowshop} is a three-stage flow-shop scheduling problem. We now establish the relationship between Problem~\ref{schedule} and Problem~\ref{flowshop}. 

\begin{lemma}[Relationship Between Problems~\ref{schedule} and \ref{flowshop}] \label{optimalS}
Given $\mathbf t_u$ and ${\mathbf t_d}$, an optimal solution $\mathbf S^*(\mathbf t_u,\mathbf t_d)$ to Problem~\ref{flowshop} is also an optimal solution to Problem~\ref{schedule}, i.e., $T_F^*(\mathbf t_u,\mathbf t_d) =  T_F(\mathbf S^*(\mathbf t_u,\mathbf t_d), \mathbf t_u,\mathbf t_d)$.
\end{lemma}

By Lemma \ref{optimalS}, instead of solving Problem \ref{schedule}, we can focus on solving Problem \ref{flowshop}. Johnson's algorithm\cite{johnson1954optimal} can guarantee to find an optimal sequence for a three-stage flow-shop problem in the special case where:
\begin{equation} \label{john_constraint}
\min_{k \in \mathcal K} \; \{t_{u,k}\} \; \ge \; \max_{k \in \mathcal K} \; \{t_{e,k}\}.
\end{equation}
In our case, \eqref{john_constraint} usually holds, as executing duration for each task at the BS is usually small due to the strong computing capability at the MEC servers. Thus, we use Johnson's algorithm to solve Problem~\ref{flowshop} approximately. If \eqref{john_constraint} holds, the obtained solution is optimal; otherwise, it is usually a sub-optimal solution with good performance.

However, even though we can efficiently solve Problem \ref{schedule}, we cannot find a simple closed-form expression for $T_F^*(\mathbf t_u,\mathbf t_d)$. Thus, it is difficult to solve Problem \ref{tran} efficiently. To reduce the complexity for solving Problem~\ref{tran}, we first neglect the BS executing duration of each task, and use Lemma~\ref{offloading_problem} to obtain the optimal uploading and downloading duration allocation with negligible BS execution duration, denoted as ($\mathbf t_u^\dag, \mathbf t_d^\dag$), as an approximation of the optimal solution ($\mathbf t_u^*, \mathbf t_d^*$) to Problem~\ref{tran}. Then, under ($\mathbf t_u^\dag, \mathbf t_d^\dag$), we solve Problem \ref{flowshop} by Johnson's algorithm to obtain a sub-optimal sequence $\mathbf S^{\dag}(\mathbf t_u^\dag,\mathbf t_d^\dag)$. 
We have the following theorem.

\begin{theorem}[Optimality of ($\mathbf t_u^\dag,\mathbf t_d^\dag$)]\label{optimal_theorem}
If $T_F(\mathbf S^\dag(\mathbf t_u^\dag,\mathbf t_d^\dag), \mathbf t_u^\dag,\mathbf t_d^\dag) \leq T$, then $(\mathbf t_u^\dag,\mathbf t_d^\dag)$ is optimal to Problem~\ref{tran}.
\end{theorem}

Note that $T_F(\mathbf S^\dag(\mathbf t_u^\dag,\mathbf t_d^\dag), \mathbf t_u^\dag,\mathbf t_d^\dag) \leq T$ indicates that the executing operations of all tasks can be conducted within the uploading and downloading durations, and hence do not take extra time.

In the worst case, $T_F(\mathbf S^\dag(\mathbf t_u^\dag,\mathbf t_d^\dag), \mathbf t_u^\dag,\mathbf t_d^\dag)$ just slightly exceeds $T$, as the BS executing duration of each task is small.
Thus, we can infer that $\mathbf S^\dag(\mathbf t_u^\dag,\mathbf t_d^\dag)$ is close to the optimal sequence $\mathbf S^*(\mathbf t_u^*$, $\mathbf t_d^*)$. Under $\mathbf S^\dag(\mathbf t_u^\dag,\mathbf t_d^\dag)$, we solve Problem \ref{C_transed} to obtain $(\mathbf t_u^*(\mathbf S^\dag(\mathbf t_u^\dag,\mathbf t_d^\dag)),\mathbf t_d^*(\mathbf S^\dag(\mathbf t_u^\dag,\mathbf t_d^\dag)))$ as an approximation of $(\mathbf t_u^*,\mathbf t_d^*)$. 
Therefore,  $(\mathbf t_u^*(\mathbf S^\dag(\mathbf t_u^\dag,\mathbf t_d^\dag)),\mathbf t_d^*(\mathbf S^\dag(\mathbf t_u^\dag,\mathbf t_d^\dag)))$ serves as a sub-optimal solution to Problem \ref{tran}. 
The details for obtaining this sub-optimal solution are summarized in Algorithm \ref{C_heuristic}.

\begin{algorithm}[h]
\caption{\small{: Sub-optimal Solution to Problem \ref{tran}}}
\small{
\begin{algorithmic}[1]
\STATE Calculate $\mathbf t_u^\dag,\mathbf t_d^\dag$ by Lemma \ref{solve_t}.
\STATE Treat  $\mathbf t_u^\dag,\mathbf t_e$ and $\mathbf t_d^\dag$ as the processing times on the three machines and use Johnson's algorithm to obtain $\mathbf S^\dag(\mathbf t_u^\dag,\mathbf t_d^\dag)$.

\IF{$T_F(\mathbf S^\dag, \mathbf t_u^\dag,\mathbf t_d^\dag) \leq T$} \STATE
$(\mathbf t_u^\dag,\mathbf t_d^\dag)$ is optimal to Problem \ref{tran}

\ELSE \STATE Obtain a sub-optimal solution $(\mathbf t_u^*(\mathbf S^\dag(\mathbf t_u^\dag,\mathbf t_d^\dag)),\mathbf t_d^*(\mathbf S^\dag(\mathbf t_u^\dag,\mathbf t_d^\dag)))$ by solving Problem \ref{C_transed} under $\mathbf S^\dag(\mathbf t_u^\dag,\mathbf t_d^\dag)$
\ENDIF
\end{algorithmic}}
\label{C_heuristic}
\end{algorithm}

\subsection{Comparison Between Optimal and Sub-optimal Solutions}
\begin{figure}[H]
\begin{center}
  \subfigure[\small{Number of users $K$ versus the overall weighted sum energy consumption.}]
  {\resizebox{4.4cm}{!}{\includegraphics{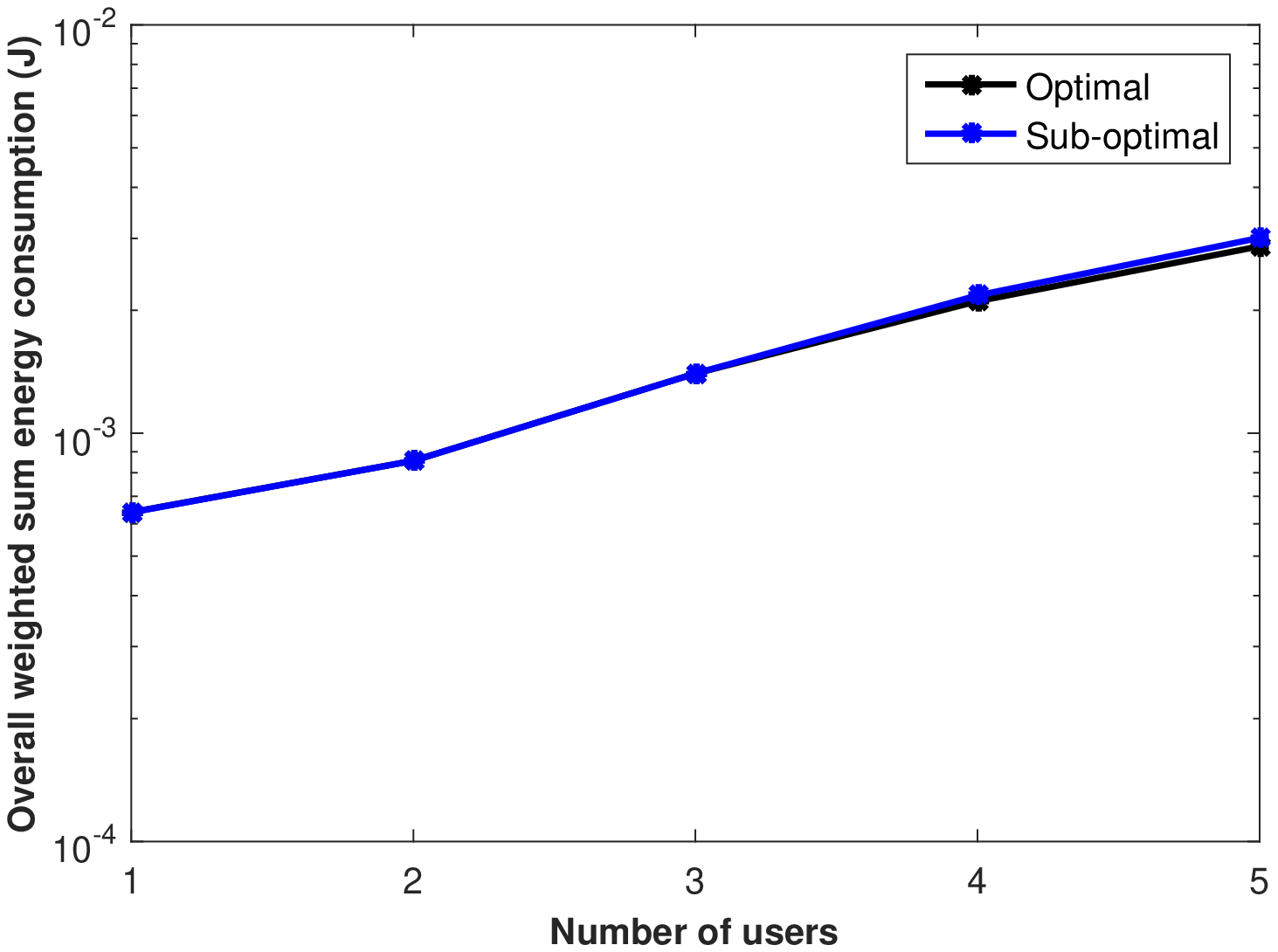}}}
  \subfigure[\small{Number of users $K$ versus the computation time.}]
  {\resizebox{4.4cm}{!}{\includegraphics{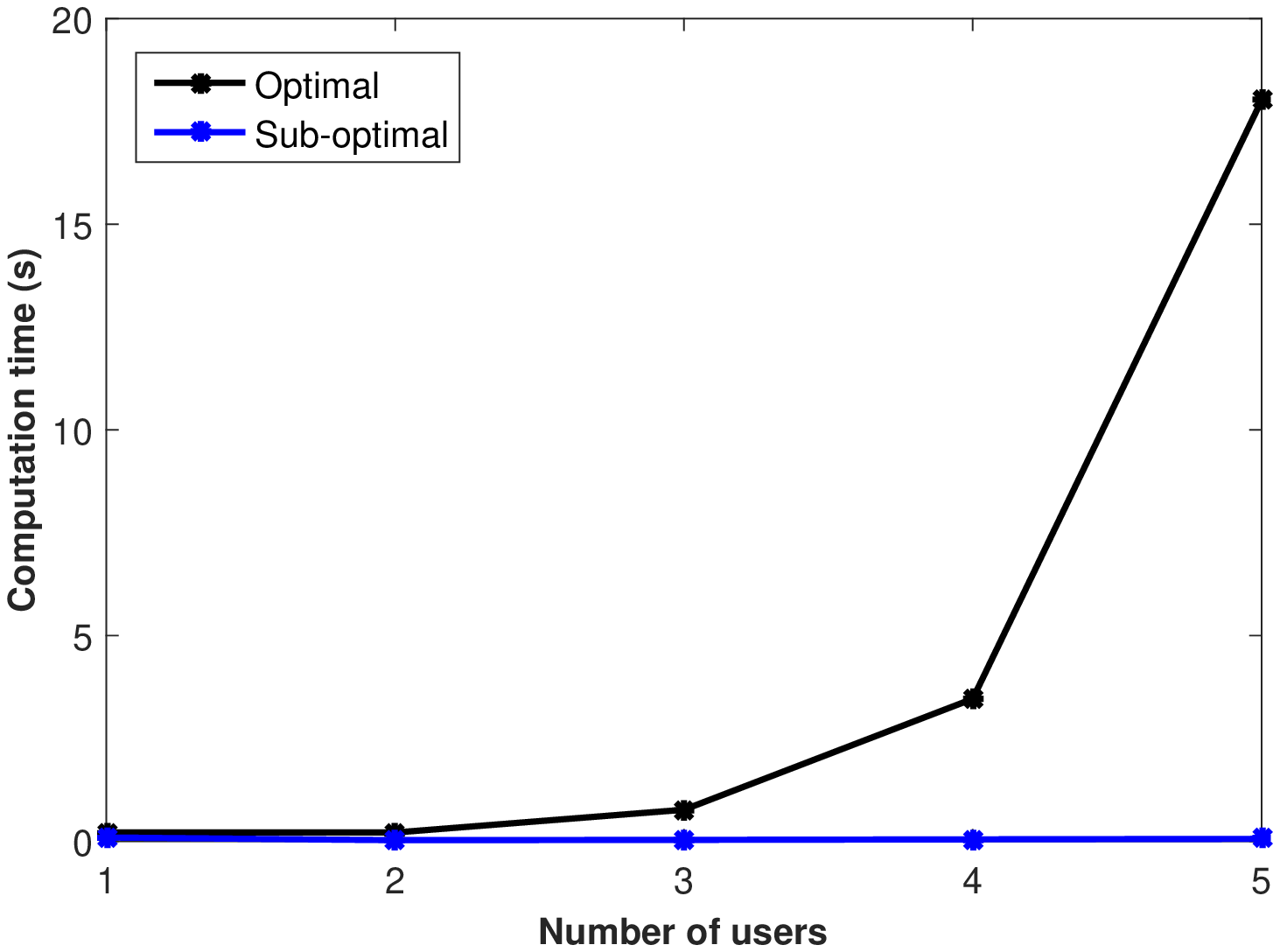}}}
  \end{center}
         \caption{\small{ Comparison between optimal and sub-optimal solutions.  }}
\label{non_neg_small}
\end{figure}

Now, we use a numerical example to compare the optimal solution and the proposed sub-optimal solution in both overall weighted sum energy consumption and computational complexity.
From Fig.~\ref{non_neg_small}(a), we can see that the performance of the proposed sub-optimal solution is very close to that of the optimal solution. From Fig.~\ref{non_neg_small}(b), we can see that the computation time for computing the sub-optimal solution grows at a much smaller rate than the optimal solution with respect to the number of users.
This numerical example demonstrates the applicability and efficiency of the sub-optimal solution.

\section{Simulation Results}
In this section, we show the performance of the proposed optimal and sub-optimal solutions for the multi-user MEC system with negligible and non-negligible BS executing durations using numerical results. Similar to \cite{you2016energy0} and \cite{you2016energy}, we consider the following simulation settings. We let $\beta =0.1$, $T = 80$ms, $\mu = 10^{-29}$, and $F=6\times10^9$. Channel power gain $h_k$ for mobile $k$ is modeled as Rayleigh fading with average power loss $10^{-3}$. The variance of complex white Gaussian channel noise is $n_0 = 10^{-9}$ W. For each task $k$, $L_{u,k}$ and  $L_{d,k}$ follow the uniform distribution over $[ 1\times10^5, 5\times10^5]$ (bits), and $N_k$ follows the uniform distribution over $[ 0.5\times10^7, 1.5\times10^7 ]$ (CPU-cycles). All random variables are independent.

\subsection{Multi-user MEC with Negligible BS Executing Duration}
In this part, we consider the multi-user MEC system with negligible BS executing duration.
We compare the proposed optimal solution (given in Lemma~\ref{offloading_problem}) with a baseline policy. The baseline policy allocates the total time $T$ equally to the uploading and downloading operations of all tasks\cite{you2016energy0,you2016energy}, i.e., $t_{u,k} = t_{u,k} = \frac{T}{2K}$ for all $k \in \mathcal K$.

\begin{figure}[t]
\begin{center}
  \subfigure[\small{Number of users $K$ at $T = 80$ms.}]
  {\resizebox{4.4cm}{!}{\includegraphics{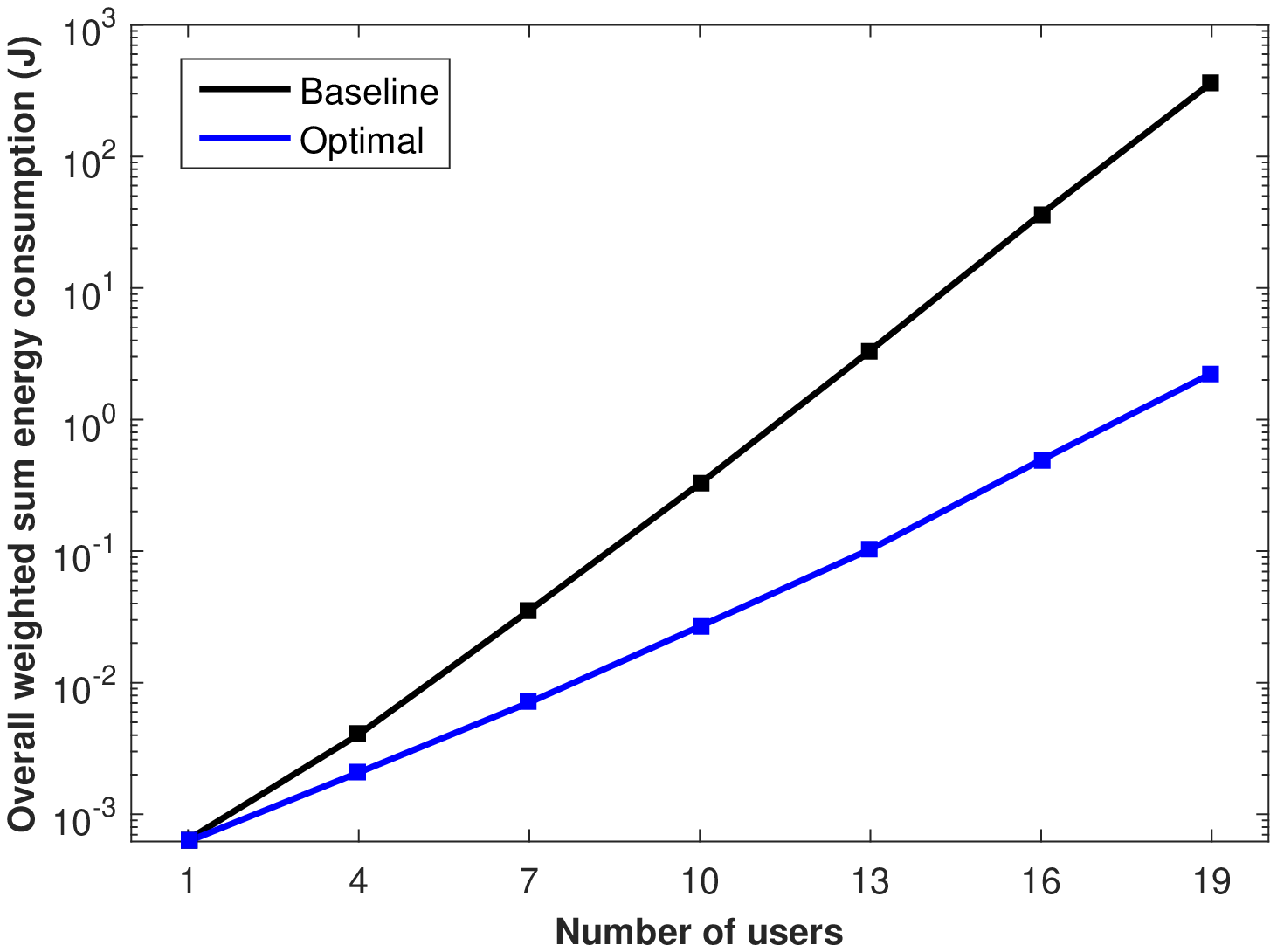}}}
  \subfigure[\small{Time duration $T$ at $K = 10$.}]
  {\resizebox{4.4cm}{!}{\includegraphics{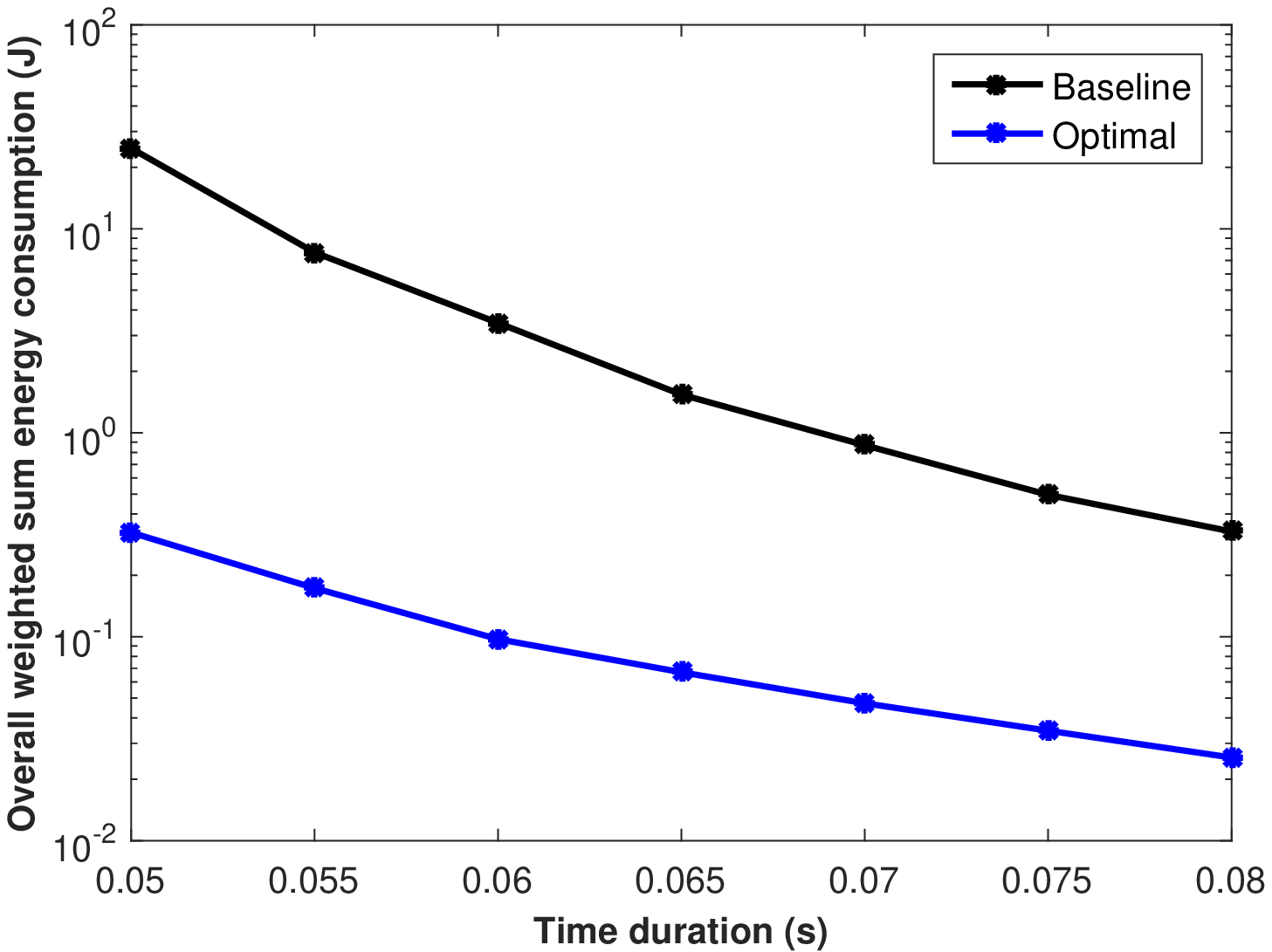}}}
  \end{center}
         \caption{\small{
         The overall weighted energy consumption versus the number of users and the time duration for the multi-user MEC system with negligible BS executing duration.
         }}
\label{big}
\end{figure}

Fig.~\ref{big}(a) and Fig.~\ref{big}(b) illustrate the overall weighted sum energy consumption versus the number of users $K$ and the time duration $T$, for the optimal solution and the baseline policy. From Fig.~\ref{big}(a) and Fig.~\ref{big}(b), we can observe that as the number of users increases or the time duration decreases, the overall weighted sum energy consumption increases. The optimal solution significantly outperforms the baseline policy, as it can optimally make use of task and channel information in reducing the overall weighted sum energy consumption.


%

\subsection{Multi-user MEC with Non-negligible BS Executing Duration}

In this part, we compare the proposed sub-optimal solution (using Algorithm \ref{C_heuristic}) with two baseline policies. Both baseline policies assume that the transmission and execution durations cannot be paralleled, and consider $T - \sum_{k=1}^{K}t_{e,k}$ as the total transmission time $\sum_{k=1}^{K}(t_{u,k}+t_{d,k})$. In particular, Baseline 1 allocates the total transmission time $T-\sum_{k=1}^{K}t_{e,k}$ equally to the uploading and downloading operations of all tasks, i.e., $t_{u,k} = t_{u,k} = \frac{T-\sum_{k=1}^{K}t_{e,k}}{2K}$ for all $k \in \mathcal K$\cite{you2016energy0,you2016energy}. Baseline 2 optimally allocates the total time to uploading and downloading operations to minimize the overall weighted sum energy consumption, using Lemma \ref{offloading_problem}.

\begin{figure}[t]
\begin{center}
  \subfigure[\small{Number of users $K$ at $T = 80$ms.}]
  {\resizebox{4.4cm}{!}{\includegraphics{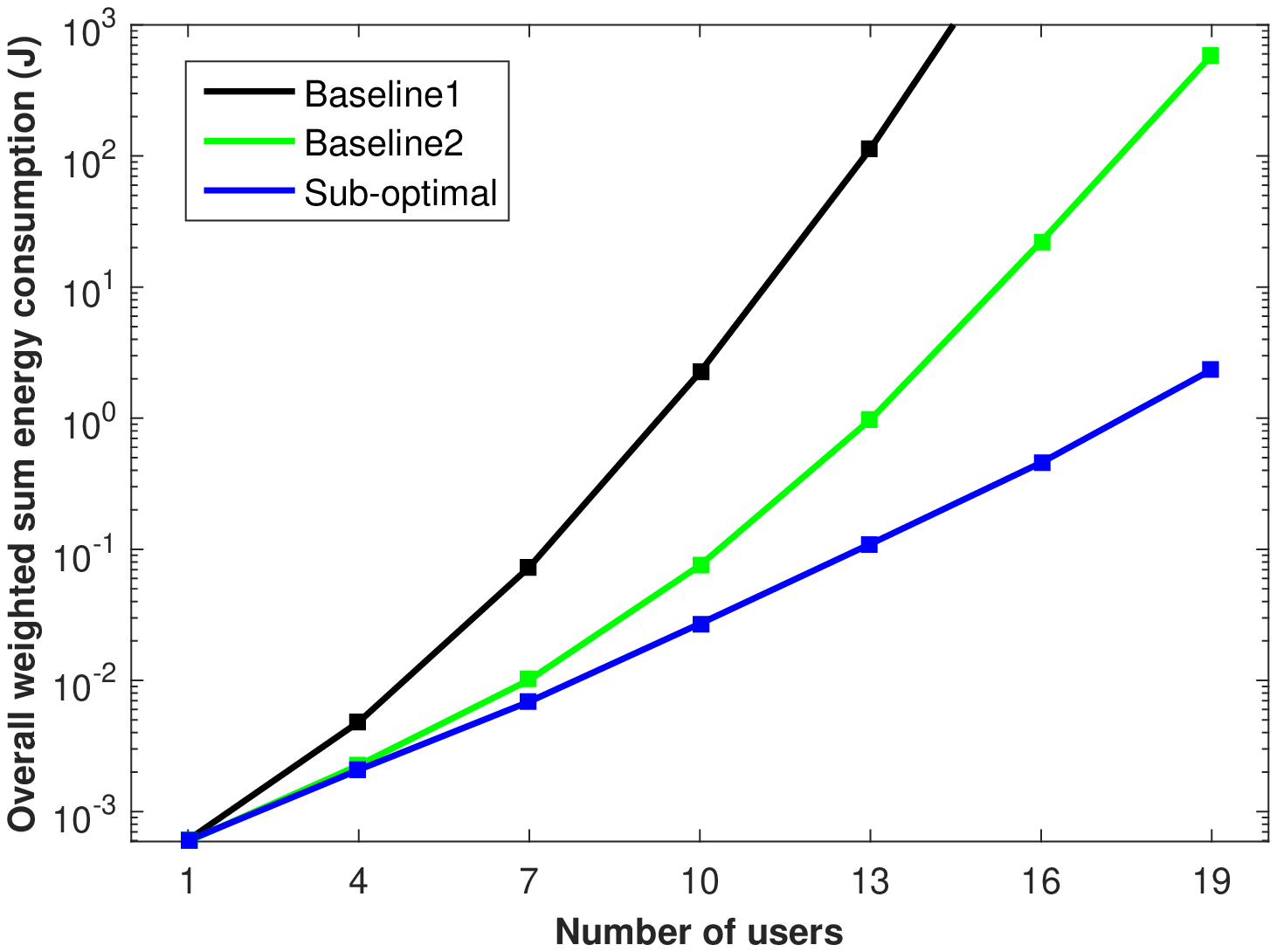}}}
  \subfigure[\small{Time duration $T$ at $K = 10$.}]
  {\resizebox{4.4cm}{!}{\includegraphics{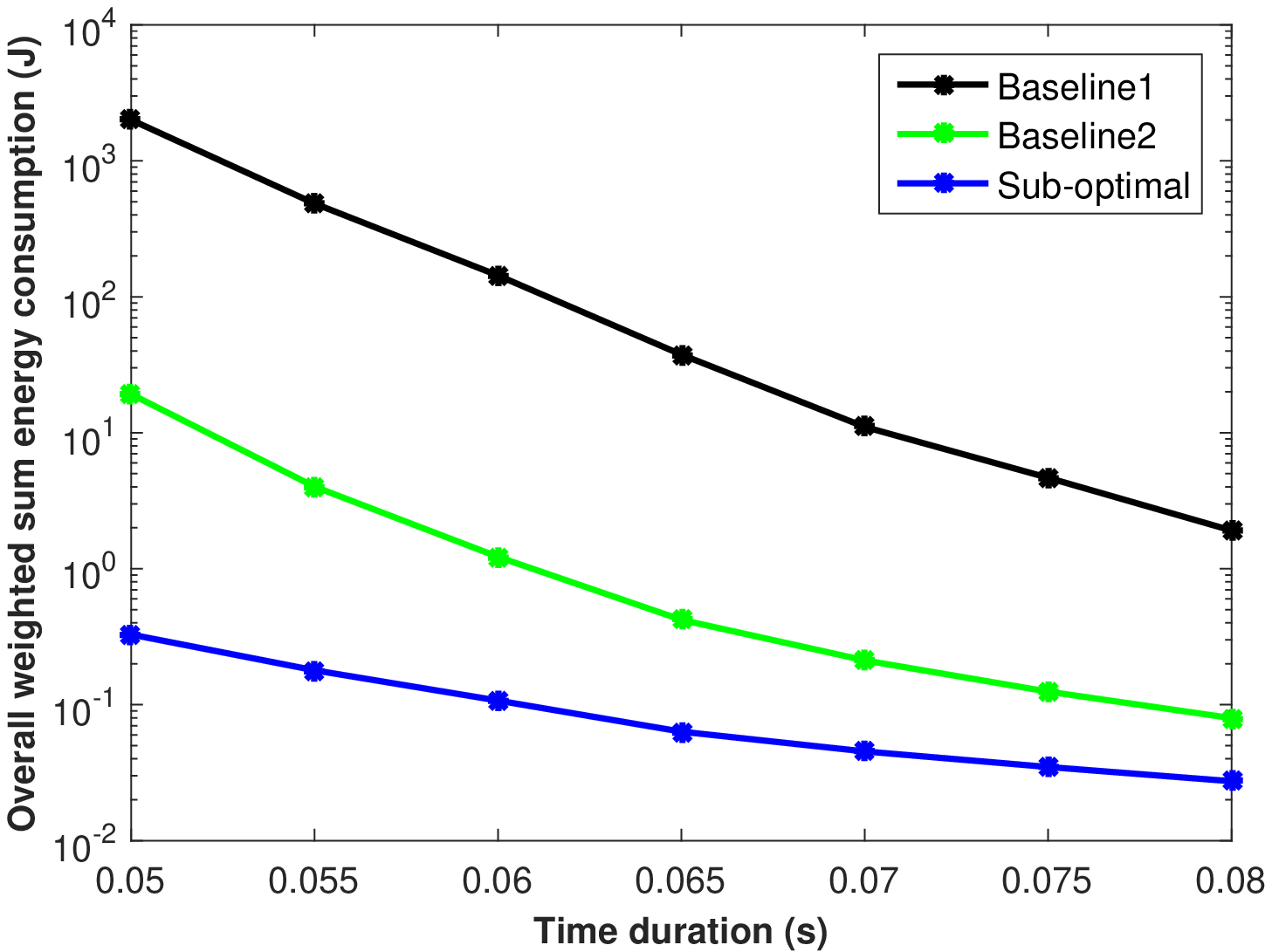}}}
  \end{center}
         \caption{\small{
         The overall weighted energy consumption versus the number of users and the time duration for the multi-user MEC system with non-negligible BS executing duration.
         }}
\label{big2}
\end{figure}

Fig.~\ref{big2}(a) and Fig.~\ref{big2}(b) illustrate the overall weighted sum energy consumption versus the number of users and the time duration, for the sub-optimal solution and the baseline policies. From Fig.~\ref{big2}(a) and Fig.~\ref{big2}(b), we can observe that as the number of users increases or the time duration decreases, the overall weighted sum energy consumption increases. The sub-optimal solution greatly outperforms Baseline 2, as it approximately maximizes the time duration over which the transmission and execution are conducted in parallel, hence maximizes the total transmission time. The sub-optimal solution significantly outperforms the baseline policies.

\section{Conclusion}
In this paper, we consider energy-efficient resource allocation for a multi-user mobile edge computing system. First, we establish on two computation-offloading models with negligible and non-negligible BS executing durations, respectively. Then, under each model, we formulate the overall weighted sum energy consumption minimization problem. The optimization problem for negligible BS executing duration is convex, and we obtain the closed-form optimal solution for each task. The optimization problem for non-negligible BS executing duration is NP-hard in general, and we obtain a low-complexity sub-optimal solution, by connecting the problem to a three-stage flow-shop scheduling problem and wisely utilizing Johnson's algorithm. Finally, numerical results show that the proposed solutions outperform some baseline schemes.

\end{document}